\documentclass[12pt]{article}

\usepackage[utf8]{inputenc}

\usepackage{amsmath,amsthm,amssymb,amscd}
\usepackage{a4wide}
\usepackage[vcentermath]{youngtab}
\usepackage{dsfont}

\usepackage{url}

\usepackage[dvips]{graphicx}
\usepackage{psfrag}
\usepackage[bf,small]{caption}
\usepackage{enumitem}

\usepackage[super,comma]{natbib}

\usepackage{sectsty}
\sectionfont{\bf\large}
\subsectionfont{\bf\normalsize}

\usepackage{color}

\definecolor{grey}{gray}{0.4}

\definecolor{mygr}{rgb}{0,0.4,0.4}

\usepackage{def}

\begin{document}

\thispagestyle{plain}

\noindent
\hspace*{1ex} \ \hfill {\tt LU-TP 13-26}
\\

\noindent
{\large\bf Hermitian Young Operators}\\[2ex]
{\bf Stefan Keppeler}${}^a$ and {\bf Malin Sjödahl}${}^b$\\[2ex]
${}^a$Mathematisches Institut, 
Universität Tübingen, 
Auf der Morgenstelle 10, 
72076 Tübingen, 
Germany, 
\url{stefan.keppeler@uni-tuebingen.de}\\
${}^b$Department of Astronomy and Theoretical Physics, 
Lunds Universitet, 
Sölvegatan 14A, 
223\,62 Lund, 
Sweden, 
\url{malin.sjodahl@thep.lu.se}\\[2ex]

\parbox{.8\textwidth}{\raggedright{\bf Abstract.} Starting from
  conventional Young operators we construct Hermitian operators which
  project orthogonally onto irreducible representations of the
  (special) unitary group.}


\section{Introduction}

Young operators play a crucial role in the representation theories of
the symmetric group $S_n$ as well as of $\GL(N)$ and the classical
compact Lie groups. They first appear when decomposing the group
algebra $\alg(S_n)$ of the symmetric group $S_n$ into a direct sum of
minimal (left) ideals, thus completely reducing the regular
reppresentation and constructing all irreducible representions. The
Young operators are primitive idempotents generating these
ideals. They can be written down in a straightforward way from the
corresponding (standard) Young tableaux.

When reducing product representations of $\GL(N)$ or classical compact
Lie groups, Young operators appear again as projectors onto
irreducible subspaces, allowing, e.g., the construction of all
irreducible representations of classical Lie groups like $\U(N)$ or
$\SU(N)$. These topics are covered by many classical text
books.\cite{Wey46,Lit50,Ham62,Sch76,Tun85}

Young operators can be cast in a diagrammatic form, resembling Feynman
diagrams, which is particularly handy in applications to general
relativity and non-abelian quantum field theories.\cite{Cvi76} The
first treatment of this kind, known to us, is by Penrose.\cite{Pen71}
More elaborate accounts, which we will also refer to in the following,
can be found in Refs.~\citenum{ElvCviKen05}
\&\citenum{Cvi08}. Following Cvitanović\cite{Cvi08} we refer to
expressions in this diagrammatic notation as birdtracks.

In applications to quantum chromodynamics (QCD) a Young operator in
the group algebra of $S_n$ can be used in order to project from (the
color part of) the Hilbert space for $n$ quarks (or for $n$
anti-quarks) onto an irreducible subspace invariant under $\SU(3)$. A
conventional Young operator, however, is in general not Hermitian, and
thus this projection is not orthogonal. But in applications to QCD
orthogonality is a desirable property.

In QCD the color of individual quarks or gluons is never observed as
they are confined into hadrons, which are color singlets. QCD color
space may therefore always be treated by summing over the color
degrees of freedom of all incoming and outgoing
particles. Consequently, it is useful to expand scattering amplitudes
into a basis of overall color singlet states. Projection operators
onto irreducible subspaces are color singlets. Thus, they are a
convenient starting set of states which one would like to extend to a
basis. If one does so, say for a process with $n$ incoming and $n$
outgoing quarks, using conventional (i.e.\ non-Hermitian) Young
operators, the resulting basis will in general be non-orthogonal --
which is a serious drawback for explicit calculations. If, however,
Hermitian Young operators are used then the resulting multiplet basis
is orthogonal, see, e.g., Appendix~B of Ref.~\citenum{CviLauSch81} or
Secs.~2~\&~3 of Ref.~\citenum{KepSjo12} for the case of 3 quarks going
to 3 quarks. Therefore, one would like to replace conventional Young
operators with Hermitian alternatives.

Hermitian Young operators for up two 3 quarks, i.e.\ for $n\leq3$, are
given in Refs.~\citenum{Can78} \& \citenum{CviLauSch81}, in
diagrammatic notation. Canning\cite{Can78} also sketches (without
proof) the idea underlying the general construction to be discussed
below. Cvitanović\cite{Cvi08} describes a general algorithm for
constructing Hermitian Young operators, and explicitly gives all
birdtrack diagrams for complete sets of Hermitian Young operators with
$n\leq4$, see Fig.~9.1 in Ref.~\citenum{Cvi08}. His method requires
the solutions to certain characteristic equations, which tend to
become more complicated for larger $n$ -- whereas conventional Young
operators always can be constructed directly from the Young tableaux.

Our main result, as summarized in Theorem~\ref{thm:Hermitian_Young}
below, is a recursive algorithm for directly constructing a Hermitian
Young operator corresponding to any given standard Young tableau. This
allows for a complete decomposition of the $n$-quark Hilbert space
into an orthogonal sum of irreducible $\SU(N)$-invariant subspaces.

The article is organized as follows. We review some properties of the
group algebra of finite groups in Sec.~\ref{sec:group_algebra} and of
Young operators in Sec.~\ref{sec:Young_operators}. In
Sec.~\ref{sec:tensors_birdtracks} we discuss product representations
of $\SU(N)$ whereby we introduce the diagrammatic birdtrack notation
and recall how the dimensions of irreducible $\SU(N)$ representations
corresponding to standard Young tableaux are calculated
diagrammatically. Our main results are stated and proven in
Sec.~\ref{sec:Hermitian_Young}, which we conclude with some
examples. Some detailed birdtrack calculations have been moved to the
Appendix.

\section{The group algebra of a finite group}
\label{sec:group_algebra}

We briefly recount some basic properties of the group algebra of a
finite group in order to fix our notation; details can by found in
standard text books.\cite{Lit50,Ham62,Sch76,Tun85}

For a finite group $G$ we define its group algebra $\alg(G)$ as the
$\C$-vector space spanned by the group elements with multiplication
induced from the group multiplication. The group algebra carries the
regular representation of the group, which contains all irreducible
representations. Minimal left ideals of $\alg(G)$ are also irreducible
$G$-invariant subspaces, and hence carry the irreducible
representations of $G$. Left ideals are generated by right
multiplication with idempotents $e\in\alg(G)$. Primitive idempotents
generate minimal left ideals. Below we will make use of the following
two statements.
\begin{lemma}\label{lemma:primitive}
  An idempotent $e$ is primitive if and only if $\forall\, r \in
  \mathcal{A}(G)$ there exists $\lambda_r \in \C$ such that
  $ere=\lambda_r e$.
\end{lemma}
\begin{lemma}\label{lemma:inequivalent}
  Two primitive idempotents $e_1$ and $e_2$ generate equivalent
  irreducible representations if and only if there exists an $r \in
  \mathcal{A}(G)$ such that $e_1 r e_2 \neq 0$.
\end{lemma}
\noindent 
Proofs can be found, e.g., in App.~III of Ref.~\citenum{Tun85}.  With
a set of primitive idempotents $e_j$, satisfying $e_j e_k =
\delta_{jk} e_j$ and $\sum_j e_j = \eins$, the regular representation
can be reduced completely, and all irreducible representations of
$G$ can be constructed. For $G=S_n$, the symmetric group, this reduction
of the regular representation is achieved in terms of Young operators.


\section{Young operators}
\label{sec:Young_operators}

Young diagrams, Young tableaux and Young operators and their
properties are discussed in many excellent text
books.\cite{Wey46,Lit50,Ham62,Sch76,Tun85,Cvi08} Nevertheless -- as scope,
conventions, and notation vary considerably between different
presentations -- we find it convenient to summarize a few definitions
and results in order to keep the presentation reasonably self-contained
and to fix our notation.

A Young diagram is an arrangement of $n$ boxes in $r$ rows of lengths
$\lambda_1 \geq \lambda_2 \geq \hdots \geq \lambda_r$. We also denote
by $s=\lambda_1$ the number of columns of the diagram, and by $\mu_1
\geq \mu_2 \geq \ldots \geq \mu_s$ the lengths of its columns. A Young
tableau $\YTab$ is a Young diagram with each of the numbers
$1,\hdots,n$ written into one of its boxes. For standard Young
tableaux the numbers increase within each row from left to right and
within each column from top to bottom. We denote the set of all
standard Young tableaux with $n$ boxes by $\SYTx_n$, e.g.
\begin{equation}
  \SYTx_2 = \left\{ \scyoung{12} \, , \ \scyoung{1,2} \right\} \, , \quad  
  \SYTx_3 = \left\{ \scyoung{123} \, , \ \scyoung{12,3} \, , \ 
                    \scyoung{13,2} \, , \ \scyoung{1,2,3} \right\} \, .
\end{equation}
Removing the box containing the number $n$ from $\YTab \in \SYTx_n$
one obtains a standard tableau $\YTab' \in \SYTx_{n-1}$.

For $\YTab\in\SYTx_n$ let $\{h_\YTab\}$ be the set of all horizontal
permutations, i.e. $h_\YTab \in S_n$ leaves the sets of numbers
appearing in the same row of $\YTab$ invariant. Analogously, vertical
permutations $v_\YTab$ leave the sets of numbers appearing
in the same column of $\YTab$ invariant. Then the Young operator $\YOp_\YTab$
is defined in terms of the row symmetrizer, $s_\YTab =
\sum_{\{h_\YTab\}} h_\YTab$, and the column anti-symmetrizer, $a_\YTab
= \sum_{\{v_\YTab\}} \mathrm{sign}(v_\YTab) v_\YTab$, as
\begin{equation}
\label{eq:definition_Y}
  \YOp_\YTab = \tfrac{1}{|\YTab|} s_\YTab a_\YTab \, .
\end{equation}
The normalization factor is given by the product of hook lengths of
the boxes of $\YTab$,
\begin{equation}
  |\YTab| = \prod_{j=1}^r \prod_{k=1}^{\lambda_j} (\lambda_j-k+\mu_k-j+1) \, .
\end{equation}
For two Young tableaux $\YTab \neq \vYTab$ of the same shape we have
$|\YTab|=|\vYTab|$, i.e. the normalization depends only on the
corresponding Young diagram. For instance, writing the hook lengths
into the boxes of $\Yboxdim{6pt}\yng(3,2)$ we obtain
\begin{equation}
  \scyoung{431,21} 
  \qquad \text{and thus} \qquad
  \Yboxdim{6pt}
  \left| \yng(3,2) \right| = 24 \, .
\end{equation}

For $\YTab \in \SYTx_n$ the corresponding Young operator $\YOp_\YTab
\in \alg(S_n)$ is a primitive idempotent. Different Young operators
$\YOp_\YTab,\YOp_{\vYTab} \in \alg(S_n)$ satisfy $\YOp_\YTab
\YOp_{\vYTab} = 0$ if the corresponding Young tableaux have different
shapes. For small $n$ one even has 
\begin{equation}\label{eq:transversality_conv_Young}
  \YOp_\YTab \YOp_{\vYTab} = \delta_{\YTab\vYTab} \YOp_\YTab
  \quad \forall\ \YTab,\vYTab \in \SYTx_n 
  \text{ and } \forall\ n \leq 4 \, , 
\end{equation}
a property to which we refer as {\it transversality}. For $n\geq5$, however,
transversality no longer holds in general, the standard
example\cite{Lit50,Sch76} being the two 5-box diagrams
$\tyoung{123,45}$ and $\tyoung{135,24}$ for which one obtains
\begin{equation}
\label{eq:Littlewood_example}
  \YOp_{_\tyoung{135,24}} \YOp_{_\tyoung{123,45}} = 0
  \qquad \text{but} \qquad
  \YOp_{_\tyoung{123,45}} \YOp_{_\tyoung{135,24}} \neq 0 \, .
\end{equation} 
Different methods for curing this are in use. One option is to choose
a different set of (non-standard) Young tableaux such that property
\eqref{eq:transversality_conv_Young} is reestablished; for
$n=5$, see e.g.\ Sec.~II.3.6 in Ref.~\citenum{Sch76}. A different way
out, which straightforwardly extends to larger $n$, is to construct
the Young operators corresponding to standard tableaux in a certain
order and to subtract multiples of already constructed operators, such
that \eqref{eq:transversality_conv_Young} holds again, see
e.g.\ Sec.~5.4 in Ref.~\citenum{Lit50} or Sec.~II.3.6 in
Ref.~\citenum{Sch76}. The Hermitian Young operators, which we will
introduce below, automatically satisfy
\eqref{eq:transversality_conv_Young} for arbitrary $n$.


\section{Tensor products, birdtracks and invariant tensors}
\label{sec:tensors_birdtracks}

Besides their crucial role for constructing the irreducible
representations of the symmetric group $S_n$, Young operators are
equally important for the representation theory of the general linear
group $\GL(N)$ and the classical compact groups $\U(N)$, $\SU(N)$,
$\uO(N)$, $\SO(N)$ and $\Sp(N)$. Motivated by applications in QCD, 
in the following we will always speak about $\SU(N)$,
although all results hold verbatim for $\U(N)$, and similarly for
$\GL(N)$.

Let $V = \C^N$ be the carrier space of the defining representation of
$\SU(N)$, and denote by $\Vbar$ its dual. Tensor products $V^{\otimes
  n}$ carry a product representation of $\SU(N)$ and -- by permutation
of the contributions in the different factors -- a representation
$D$ of $S_n$ as well as its group algebra $\alg(S_n)$. The
same holds for tensor products $\Vbar^{\otimes n}$ of the dual.

$V$ is naturally endowed with a scalar product
$\langle\cdot,\cdot\rangle_V$ which is invariant under the action of
$\SU(N)$, i.e.
\begin{equation}
\label{eq:canonical_SP}
  \langle g v , g w \rangle_V = \langle v,w \rangle_V 
  \qquad \forall\ v,w \in V \quad \forall\ g \in \SU(N) \, .
\end{equation}
This also induces scalar products on the dual $\Vbar$ and on tensor
products. Young operators act as projectors onto $\SU(N)$-invariant
irreducible subspaces of $V^{\otimes n}$. Irreducible representations
of $\SU(N)$ corresponding to different Young diagrams are
inequivalent.
In general, however, Young operators are not Hermitian with respect to
the scalar product induced by \eqref{eq:canonical_SP} on the product
space $V^{\otimes n}$. Thus, these projections are in general not
orthogonal.

In the diagrammatic birdtrack notation, we assign to $P: V^{\otimes n}
\to V^{\otimes n}$, i.e.\ $P \in V^{\otimes n} \otimes \Vbar^{\otimes
  n}$, a diagram with $2n$ external lines, which can be translated to
index notation as follows,
\begin{equation}
  \parbox{5cm}{\includegraphics[width=5cm]{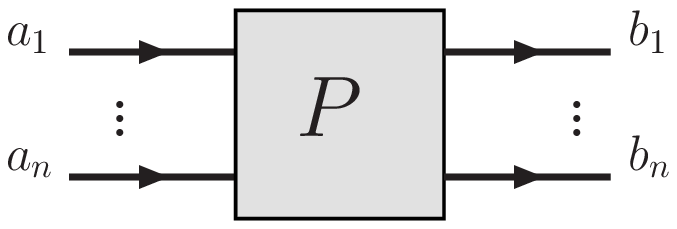}}
  = P^{a_1 \hdots a_n}{}_{b_1 \hdots b_n} \, .
\end{equation} 
Correspondingly, we represent an operator $\Vbar^{\otimes n} \to
\Vbar^{\otimes n}$ by a diagram with all arrows pointing in the
opposite direction. The diagram of the Hermitian conjugate of an
operator is obtained from its diagram by mirroring about a vertical
axis and inverting all arrows. Index contractions are performed by
joining lines. See e.g.\ App.~A of Ref.~\citenum{KepSjo12} for a brief
summary of our conventions or Cvitanović\cite{Cvi08} for a more
detailed account of the birdtrack notation.


Symmetrization or anti-symmetrization over a set of indices is
indicated by a white or black bar, respectively, e.g.\
\begin{equation}\label{eq:definition_S&A}
\begin{split} 
\parbox{1.5cm}{\includegraphics[width=1.5cm]{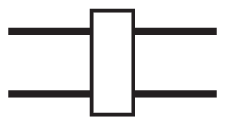}}
&= \frac{1}{2!} \left( 
   \parbox{1.2cm}{\includegraphics[width=1.2cm]{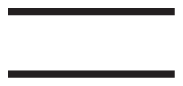}}
   + \parbox{1.4cm}{\includegraphics[width=1.4cm]{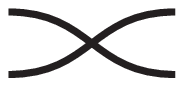}}
   \right) \, , \\
\parbox{1.5cm}{\includegraphics[width=1.5cm]{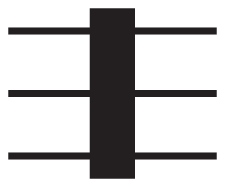}}
&= \frac{1}{3!} \left( 
   \parbox{1.2cm}{\includegraphics[width=1.2cm]{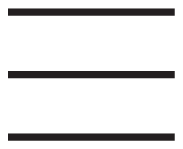}}
   - \parbox{1.4cm}{\includegraphics[width=1.4cm]{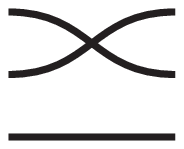}}
   - \parbox{1.4cm}{\includegraphics[width=1.4cm]{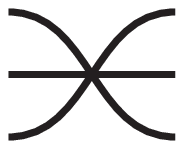}} 
   - \parbox{1.4cm}{\includegraphics[width=1.4cm]{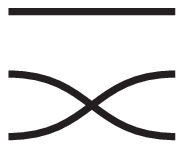}}
   + \parbox{1.4cm}{\includegraphics[width=1.4cm]{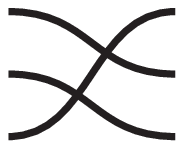}}
   + \parbox{1.4cm}{\includegraphics[width=1.4cm]{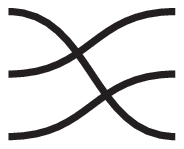}}
   \right) \, , 
\end{split} 
\end{equation}
whereby we include a factorial as normalization factor. For all
diagrams appearing in the following all arrows on external lines will
point in the same direction. We therefore omit arrows from now on.
The following recursion relations, which are derived in Chap.~6 of
Ref.~\citenum{Cvi08}, will be used below,
\begin{equation}\label{eq:recursion}
\begin{split} 
\raisebox{.3mm}{\parbox{2.95cm}{
\includegraphics[width=2.95cm]{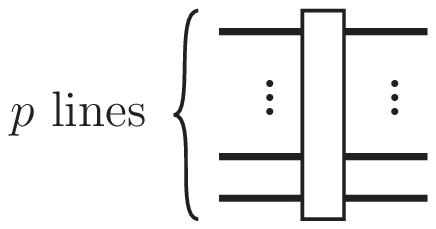}}}
\hspace*{-2ex}
&= \ \frac{1}{p} \raisebox{.8mm}{
   \parbox{1.9cm}{\includegraphics[width=1.9cm]{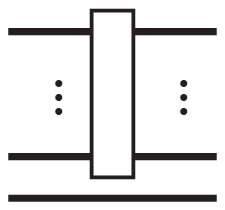}}} 
   \hspace*{-2ex} + \ \frac{p-1}{p} \raisebox{.8mm}{
   \parbox{3.1cm}{\includegraphics[width=3.1cm]{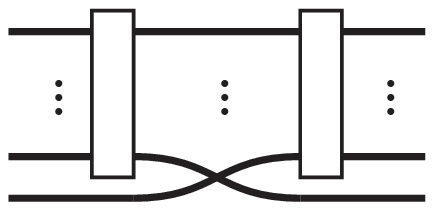}}} , 
\\ 
\parbox{2.95cm}{\includegraphics[width=2.95cm]{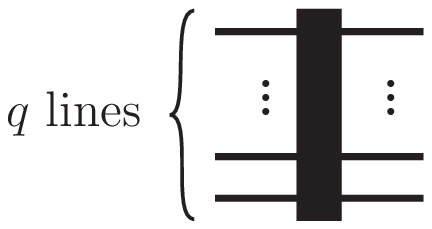}} 
\hspace*{-2ex}
&= \ \frac{1}{q} \raisebox{.8mm}{
   \parbox{1.9cm}{\includegraphics[width=1.9cm]{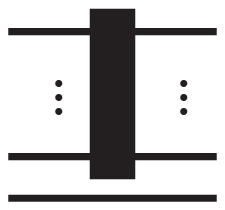}}} 
   \hspace*{-2ex} - \ \frac{q-1}{q} \raisebox{.8mm}{
   \parbox{3.1cm}{\includegraphics[width=3.1cm]{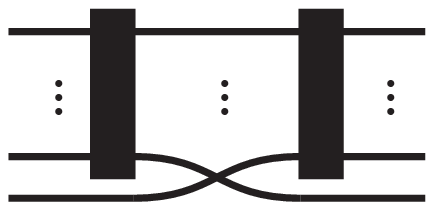}}} .
\end{split}
\end{equation}

With this notation the Young operators \eqref{eq:definition_Y} of
Sec.~\ref{sec:Young_operators} can be written
diagrammatically,\cite{ElvCviKen05,Cvi08} e.g.\ we find
\begin{equation} 
\label{eq:Littlewood_birdtracks}
  \YOp_{_\tyoung{123,45}} 
  = 2 \parbox{2.4cm}{\includegraphics[width=2.4cm]{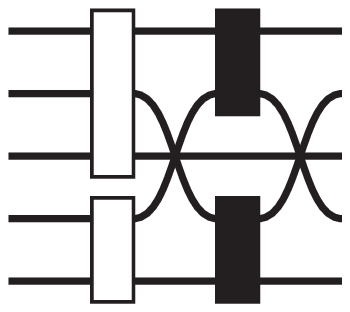}}
  \qquad \text{and} \qquad  
  \YOp_{_\tyoung{135,24}} 
  = 2 \parbox{2.4cm}{\includegraphics[width=2.4cm]{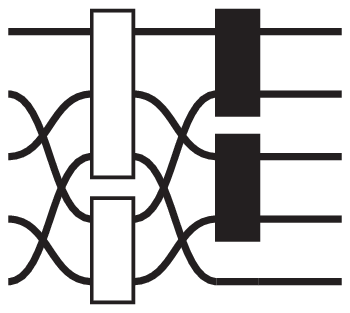}} \ . 
\end{equation} 
Notice that the pre-factor equals $3!(2!)^3/|\Yboxdim{6pt}\yng(3,2)| =
2$ in agreement with \eqref{eq:definition_Y} and the normalization
\eqref{eq:definition_S&A} of the (anti-)symmetrizers. From the
birdtrack diagrams the first equation in \eqref{eq:Littlewood_example}
is obvious, as in this product the first two lines both join the same
anti-symmetrizer and symmetrizer.

Strictly speaking, we should only write $\YOp_\YTab$ when considering
Young operators as elements of the group algebra
$\mathcal{A}(S_n)$. Viewing them as linear maps $V^{\otimes n} \to
V^{\otimes n}$ they should be denoted by $D(\YOp_\YTab)$. However, in
order to simplify our notation, we omit the $D$ in the following.

We are now in the position to state the dimension formula for the
$\SU(N)$-representations.

\begin{lemma} Let $\YTab \in \SYTx_n$. The dimension of the
  $\SU(N)$-invariant irreducible subspace of $V^{\otimes n}$ onto
  which $\YOp_\YTab$ projects is given by
  \begin{equation}
    \utr \YOp_\YTab = \frac{f_\YTab(N)}{|\YTab|}
  \end{equation}
  where 
  \begin{equation}
  \label{eq:f_YTab-Def}
    f_\YTab(N) = \prod_{j=1}^r \prod_{k=1}^{\lambda_j} (N+k-j) \, .
  \end{equation} 
\end{lemma}

\proof Elvang et al.\cite{ElvCviKen05}\ give a birdtrack proof of this
dimension formula (see also App.~B.4 of Ref.~\citenum{Cvi08}), which
we sketch here since we will use an intermediate result in the
following section.

The proof is by induction in $n$. First notice that the dimension
formula holds for $n=1$, $\Yboxdim{6pt} \utr Y_{\yng(1)}=N$. Diagrammatically,
traces are taken by joining left legs to right legs, and a loop yields
a factor $\dim V = N$. In order to establish the induction step we represent 
$\YOp_\YTab \in \SYTx_n$ in terms of the following birdtrack diagram, 
\begin{equation}
\begin{split}
  \YOp_\YTab 
  &\ = \ 
  \frac{\prod\limits_{j=1}^r \lambda_j! \prod\limits_{k=1}^s \mu_k!}{|\YTab|}
  \left. \parbox{7cm}{\includegraphics[width=7cm]{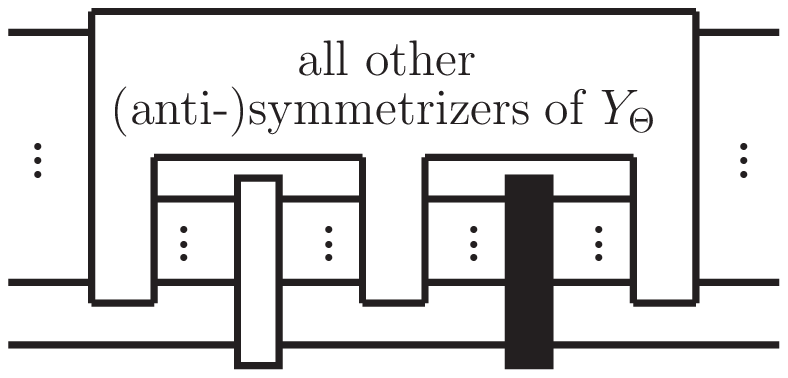}} 
  \hspace*{-3ex}\right\} n\text{ lines} \ , \\ 
  & \hspace{27ex} \uparrow p\text{ lines} 
    \hspace{7ex} \uparrow q\text{ lines} 
\end{split}
\end{equation}
in which we explicitly display the symmetrizer and anti-symmetrizer to
which the last line connects, i.e.\ the box with the highest number in
$\YTab$ is the last box of a line with $p$ boxes and of a row with $q$
boxes. All other (anti-)symmetrizers are collected in the white box.
Note that we allow for the possibilities $p=1$ or $q=1$ in which case
the respective (anti-)symmetrizer could be omitted.
 
Now take a partial trace of the last factor, denoting this operation
by $\utr':V^{\otimes n} \to V^{\otimes (n-1)}$,
\begin{equation}\label{eq:trY_01}
  \utr' \YOp_\YTab 
  = \frac{\prod\limits_{j=1}^r \lambda_j! \prod\limits_{k=1}^s \mu_k!}{|\YTab|}
    \parbox{5cm}{\includegraphics[width=5cm]{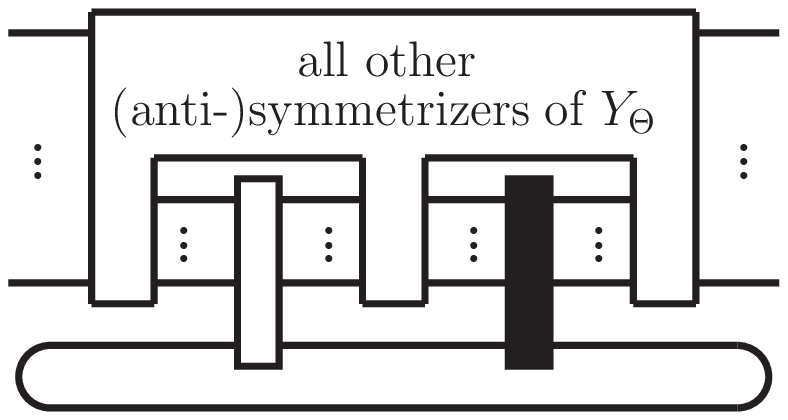}} .
\end{equation}
Using the recursion relations \eqref{eq:recursion}, the diagram in
Eq.~\eqref{eq:trY_01} can be readily reduced to
\begin{equation}\label{eq:trY_02}
\begin{split} 
  \parbox{5cm}{\includegraphics[width=5cm]{Pics/trY_01.eps}} \hspace*{-4mm}
  &= \ \frac{N+p-q}{pq} \overbrace{\raisebox{2.8mm}{
     \parbox{5cm}{\includegraphics[width=5cm]{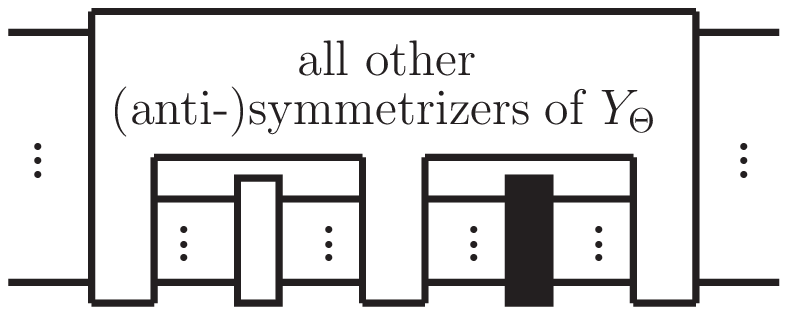}}\hspace*{-2mm}}}^{:=B_{\YTab'}}
  \\ &\qquad 
  - \frac{(p-1)(q-1)}{pq} 
    \parbox{6.5cm}{\includegraphics[width=6.5cm]{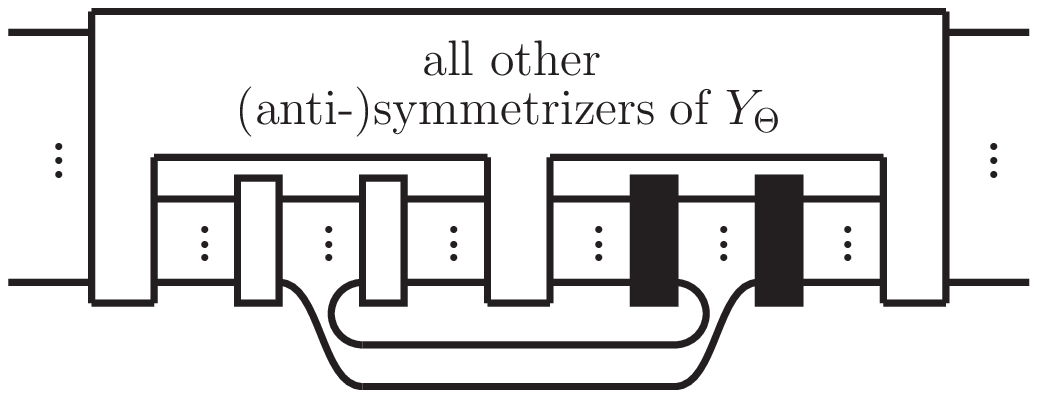}} 
  \hspace*{-3mm} , 
\end{split} 
\end{equation}
where the first term is proportional to $\YOp_{\YTab'}$, with $\YTab'$
being the Young tableau which one obtains by removing from $\YTab$ the
box with highest number. We temporarily denote the birdtrack part of
this term by $B_{\YTab'}$. Elvang et al.\
show\cite{ElvCviKen05,Cvi08}\ that the second term of
Eq.~\eqref{eq:trY_02} vanishes since on the outside all lines are
connected to (anti-)symmetrizers in the same way as in
$\YOp_{\YTab'}$, but internally the leftmost symmetrizer and the
rightmost anti-symmetrizer are connected by a line, which
automatically leads to a vanishing diagram since there can be no such
connection within $\YOp_{\YTab'}$. Collecting all contributions one
finds
\begin{equation}\label{eq:Y_recursion}
\begin{split} 
  \utr' Y_\YTab 
  &= \frac{N+p-q}{pq}\,  
     \frac{\prod\limits_{j=1}^r \lambda_j! \prod\limits_{k=1}^s \mu_k!}{|\YTab|}
     \, B_{\YTab'} 
   = (N+p-q) \frac{|\YTab'|}{|\YTab|}\, 
     \frac{\prod\limits_{j=1}^{r'} \lambda'_j! \prod\limits_{k=1}^{s'} \mu'_k!}
          {|\YTab'|}
     \, B_{\YTab'} \\
  &= (N+p-q) \frac{|\YTab'|}{|\YTab|} \YOp_{\YTab'}
\end{split}
\end{equation}
where we have used that $(\prod_{j=1}^r \lambda_j!)/p =
\prod_{j=1}^{r'} \lambda'_j!$ since the row lengths $\lambda'_j$ of
$\YTab'$ are the same as those of $\YTab$ except for the row from
which one removes a box, which has length $p$ in $\YTab$ but length
$(p-1)$ in $\YTab'$. The analogous statement holds for the column
lengths and the factor $q$. Taking the trace, and observing that
$(N+p-q)$ is the contribution to $f_\YTab(N)$ in
\eqref{eq:f_YTab-Def} coming from the box which distinguishes $\YTab$
from $\YTab'$, we get
\begin{equation}
  \utr Y_\YTab 
  = (N+p-q) \frac{|\YTab'|}{|\YTab|} \utr \YOp_{\YTab'}
  = (N+p-q) \frac{|\YTab'|}{|\YTab|} \frac{f_{\YTab'}(N)}{|\YTab'|} 
  = \frac{f_{\YTab}(N)}{|\YTab|} \, ,
\end{equation}
which concludes the induction step.\hfill \qed


We continue this section by discussing properties of invariant
tensors\cite{Cvi08} following from Schur's lemma.

\definition (Invariant tensor)\\
Let $W_1$ and $W_2$ be vector spaces carrying representations
$\Gamma_1$ and $\Gamma_2$ of $\SU(N)$ and let $T: W_1 \to W_2$ be
linear. $T$ is called {\it invariant tensor} if
\begin{equation}
  T \circ \Gamma_1(g) = \Gamma_2(g) \circ T \qquad \forall\ g \in \SU(N) \, .
\end{equation}

\Remarks
\begin{enumerate}[itemsep=0mm,topsep=0mm,parsep=0mm,partopsep=0mm]
\item Invariant tensors are defined in the same way for other compact Lie
  groups.
\item In the following we are mainly interested in the case $W_1 = W_2
  = V^{\otimes n}$ (or $\overline{V}^{\otimes n}$, leading to
  equivalent results).
\item Invariance of the scalar product \eqref{eq:canonical_SP} is equivalent
  to invariance of the tensor $\delta{^j}{_k}$, $j,k=1,\hdots,N$, (in
  index notation), or invariance of the quark
  line 
  \parbox{1.7cm}{\includegraphics[width=1.7cm]{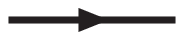}}$: V \to V$
  (in birdtrack notation).
\item Since 
\begin{equation} 
  \Gamma(g) \circ D(r) = D(r) \circ \Gamma(g)
  \qquad \forall\ g \in \SU(N) \text{ and } \forall\ r \in \mathcal{A}(S_n)
\end{equation}
every $r$ in the group algebra of $S_n$ corresponds to an invariant
tensor $D(r): V^{\otimes n} \to V^{\otimes n}$.
\end{enumerate}

\begin{lemma} {\rm (Schur)}\\
Let $T: W_1 \to W_2$ be an invariant tensor and $\Gamma_{1,2}$
irreducible representations. If $\Gamma_1$ and $\Gamma_2$ are
non-equivalent then $T=0$.  If $\Gamma_1$ and $\Gamma_2$ are
equivalent and if $W_1=W_2$ then $T$ is a multiple of the identity.
\end{lemma}

For the proof see any standard book on group and representation
theory. Simon's proof\cite{Sim96} is particularly concise. Below we
will use the following two consequences of Schur's lemma.

\begin{corollary}\label{corr:invariant_tensors_schur}
Let\ $T: W \to W$ be both, an invariant tensor and a (possibly
non-orthogonal) projection onto a subspace $U \subset W$ carrying the
irreducible representation $\Gamma$ of $\SU(N)$. 
\begin{enumerate}[itemsep=1ex,topsep=1ex,parsep=0mm,partopsep=0mm]
\item[\rm(i)] Upon decomposing $W$ into an orthogonal sum $\bigoplus_j W_j$ of
  subspaces $W_j$ carrying irreducible representations $\Gamma_j$ of
  $\SU(N)$, with $\Gamma_1,\hdots,\Gamma_m$ equivalent to $\Gamma$,
  and $\Gamma_j$, $j>m$ not equivalent to $\Gamma$, and representing
  $w \in W$ as
\begin{equation}
  w = \begin{pmatrix} w_1 \\ \vdots \\ w_m \\ w_{m+1} \\ \vdots \end{pmatrix}
  \, , \quad w_j \in W_j \, , 
\end{equation} 
  $T$ assumes the block structure 
\begin{equation}
  T = \begin{pmatrix} T_{11} & \cdots & T_{1m} & 0 & \cdots \\
                      \vdots & \ddots & \vdots & \vdots \\
                      T_{m1} & \cdots & T_{mm} &  &  \\
                      0 & \cdots & & 0  & \\
                      \vdots & & & & \ddots \end{pmatrix} \, .
\end{equation}
The diagonal blocks $T_{jj}$, $j \leq m$, are proportional to the
identity, and the $T_{jk}$ vanish if $j>m$ or $k>m$ (as already
displayed above).
\item[\rm(ii)] 
  Let, furthermore, $P: W \to W$ be a Hermitian projector, $P^\dag=P$,
  onto an invariant subspace $\widetilde{W} \subseteq W$, which
  contains only \emph{one} subspace $\widetilde{U} \subseteq
  \widetilde{W}$ carrying a representation equivalent to $\Gamma$, and
  let $PTP\neq0$. Then $PTP$ is a multiple of the orthogonal
  projection onto $\widetilde{U}$.
\end{enumerate}
\end{corollary}

\proof By applying Schur's lemma to each of the blocks $T_{jk}: W_k
\to W_j$ (i) follows immediately. Since $P$ is Hermitian its image is
orthogonal to its kernel, $\im P = (\ker P)^\perp$, i.e.\ $W$ is an
orthogonal sum, $W=\im P \oplus \ker P$. We further decompose $\im P$
into $\widetilde{U}$ and its orthogonal complement
$\widetilde{U}^\perp \subseteq \im P$. Writing $w \in W$ as
\begin{equation}
  w = \begin{pmatrix} u_1 \\ u_2 \\ u_3 \end{pmatrix} 
  \quad \text{with} \quad 
  u_1 \in \widetilde{U} \, , \ 
  u_2 \in \widetilde{U}^\perp \, , \
  u_3 \in \ker P \, ,  
\end{equation}
$P$ takes the form $P = \left( \begin{smallmatrix} \eins & 0 & 0 \\
    0 & \eins & 0 \\
    0 & 0 & 0 \end{smallmatrix} \right)$,
and according to (i) we have 
\begin{equation}
  T = \begin{pmatrix} T_{11} & 0 & T_{13} \\ 
                      0 & 0 & 0 \\ 
                      T_{31} & 0 & T_{33} \end{pmatrix} \, , 
\end{equation}
with $T_{11}$ proportional to the identity. Now by direct computation
$PTP = \left( \begin{smallmatrix} T_{11} & 0 & 0 \\
    0 & 0 & 0 \\
    0 & 0 & 0 \end{smallmatrix} \right)$, which is proportional to
$\left( \begin{smallmatrix} \eins & 0 & 0 \\
    0 & 0 & 0 \\
    0 & 0 & 0 \end{smallmatrix} \right)$, the orthogonal projection
onto $\widetilde{U}$. \hfill \qed

\section{Hermitian Young operators}
\label{sec:Hermitian_Young}

As Young operators $\YOp_\YTab$ are elements of the group algebra
$\mathcal{A}(S_n)$ they act as invariant tensors $V^{\otimes n} \to
V^{\otimes n}$. Moreover, as already mentioned, each Young operator
$\YOp_\YTab$ with $\YTab \in \SYTx_n$ projects onto an
$\SU(N)$-invariant irreducible subspace of $V^{\otimes n}$. As an
invariant tensor a Young operator cannot map an invariant subspace to
a subspace carrying a non-equivalent representation. In general,
however, a tensor product $V^{\otimes n}$ can contain several
(irreducible) subspaces carrying equivalent representations. A
projector onto such a subspace can then also contain parts mapping one
invariant subspace to another one carrying an equivalent
representation. Only through this mechanism can projectors onto
invariant subspaces be non-Hermitian. This happens with conventional
Young operators for $n \geq 3$, e.g.
\begin{equation}
  \YOp_{_\tyoung{12,3}} = \frac{4}{3} 
  \parbox{2.4cm}{\includegraphics[width=2.4cm]{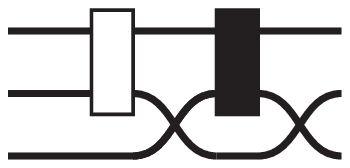}} 
  \qquad \text{and} \qquad 
  \YOp_{_\tyoung{13,2}} = \frac{4}{3} 
  \parbox{2.4cm}{\includegraphics[width=2.4cm]{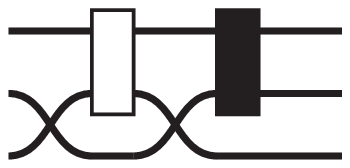}} 
\end{equation} 
are both non-Hermitian. One can, however, find the Hermitian
operators\cite{Can78,CviLauSch81,Cvi08}
\begin{equation}
  \HYOp_{_\tyoung{12,3}} = \frac{4}{3} 
  \parbox{3.0cm}{\includegraphics[width=3.0cm]{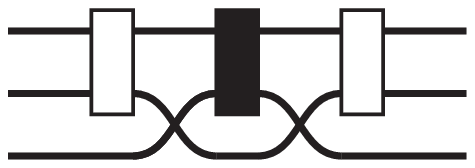}} 
  \qquad \text{and} \qquad 
  \HYOp_{_\tyoung{13,2}} = \frac{4}{3} 
  \parbox{3.0cm}{\includegraphics[width=3.0cm]{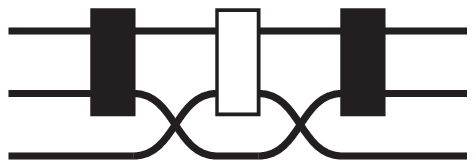}} 
\end{equation} 
which also project onto irreducible subspaces carrying the
representation corresponding to to the Young diagram
$\Yboxdim{6pt}\yng(2,1)$ and which satisfy
\begin{equation}
  \HYOp_{_\tyoung{12,3}} + \HYOp_{_\tyoung{13,2}} 
  = \YOp_{_\tyoung{12,3}} + \YOp_{_\tyoung{13,2}} \, .
\end{equation} 
In the following theorem we show how to construct, for any given
standard tableau, a Hermitian Young operator (and thus an orthogonal
projection) by suitably projecting the corresponding conventional
Young operator.

\begin{theorem}\label{thm:Hermitian_Young}
Let $\HYOp_\YTab := \YOp_\YTab$ for $\YTab \in \SYTx_2$ and 
\begin{equation} 
\label{eq:Hermitian_Young_Def}
  \HYOp_\YTab 
  := (\HYOp_{\YTab'} \otimes \eins) \YOp_\YTab (\HYOp_{\YTab'} \otimes \eins)
  \qquad \text{for} \ \YTab \in \SYTx_n \, , \quad n \geq 3 \, , 
\end{equation}
where $\YTab' \in \SYTx_{n-1}$ denotes the standard Young tableaux
obtained from $\YTab$ by removing the box with the highest number.
Then the operators $\HYOp_\YTab$ are (a complete set of) Hermitian
transversal Young projectors, i.e.\ they satisfy
\begin{enumerate}[label={\rm(\roman*)}]%
\item \label{prop:transversal} 
$\HYOp_\YTab \HYOp_{\vYTab} = \delta_{\YTab\vYTab} \HYOp_{\YTab} 
  \quad \forall\ \YTab,\vYTab \in \SYTx_n$ (transversality), 
\item \label{prop:dim} 
$\utr \HYOp_\YTab = \utr \YOp_\YTab \quad \forall\ \YTab \in \SYTx_n$
(dimension),
\item \label{prop:complete} 
$\displaystyle\sum_{\YTab \in \SYTx_n} \HYOp_\YTab = \eins^{\otimes n}$ 
(completeness) and
\item \label{prop:Hermitian}
$\HYOp_\YTab^\dag = \HYOp_\YTab$ (Hermiticity).
\end{enumerate}

\end{theorem}

\remark 
In the representation theory of the symmetric group the $\HYOp_\YTab$
are known as semi-normal idempotents \cite{You31,Thr41,Rut48} and were
introduced by Thrall\cite{Thr41} who also proved properties
\ref{prop:transversal} and \ref{prop:complete}.

\proof As the construction is recursive we prove the properties by
induction. For $n=2$ the conventional Young operators are Hermitian
and properties \ref{prop:transversal}--\ref{prop:complete} are also
fulfilled trivially.

We begin with the dimensions of the irreducible subspaces,
property~\ref{prop:dim}. Taking a partial trace of the last factor in
$\HYOp_\YTab$ and using Eq.~\eqref{eq:Y_recursion} yields
\begin{equation}
\label{eq:partial_trace_PTheta}
\begin{split} 
 \utr' \HYOp_\YTab 
 &= \parbox{3.6cm}{\includegraphics[width=3.6cm]{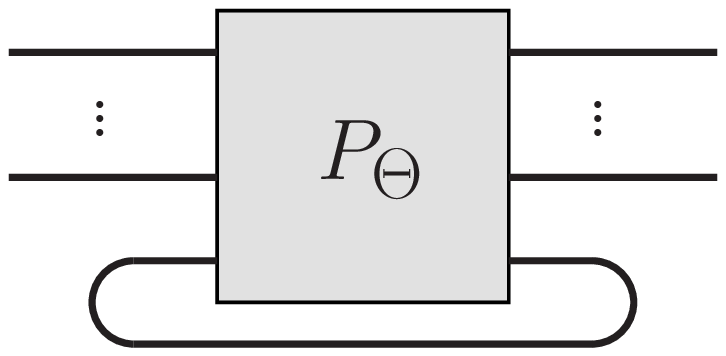}}
 = \parbox{7.5cm}{\includegraphics[width=7.5cm]{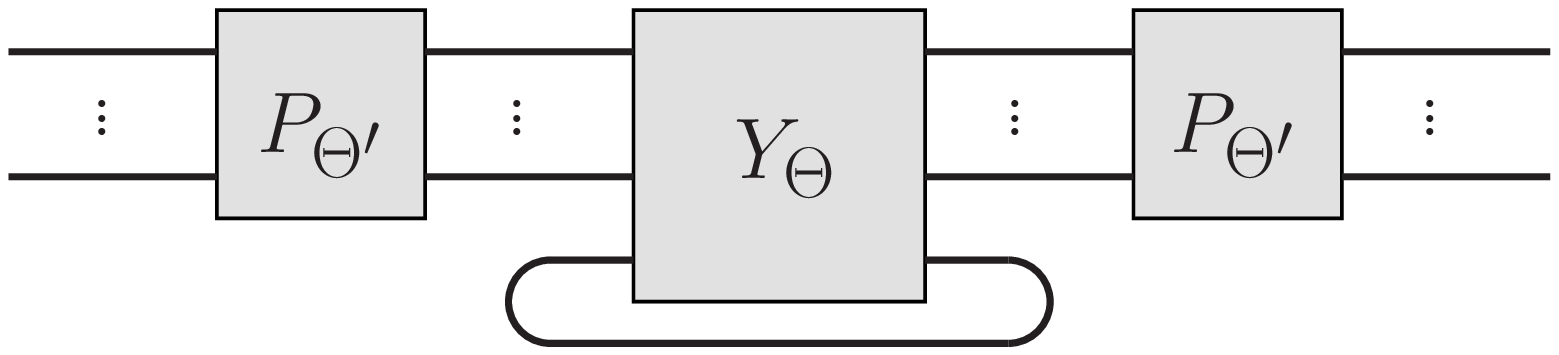}}\\[1ex]
 &= (N+p-q) \frac{|\YTab'|}{|\YTab|} 
    \parbox{7.5cm}{\includegraphics[width=7.5cm]{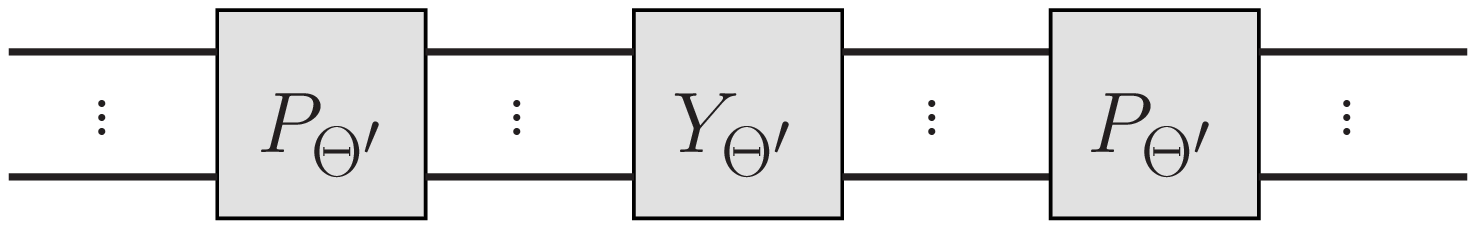}} \, .
\end{split}
\end{equation}
From the recursive definition \eqref{eq:Hermitian_Young_Def} it follows that
\begin{equation}
  \parbox{3.25cm}{\includegraphics[width=3.25cm]{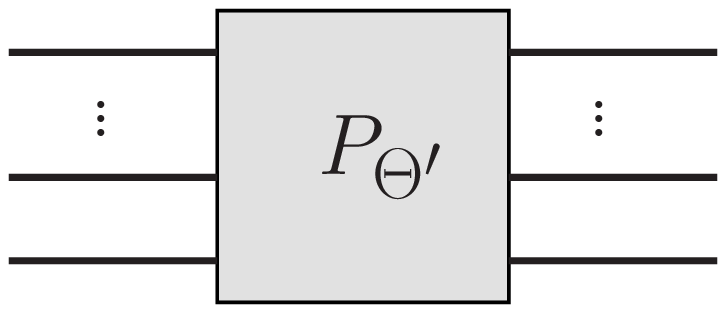}}
  = \parbox{5cm}{\includegraphics[width=5cm]{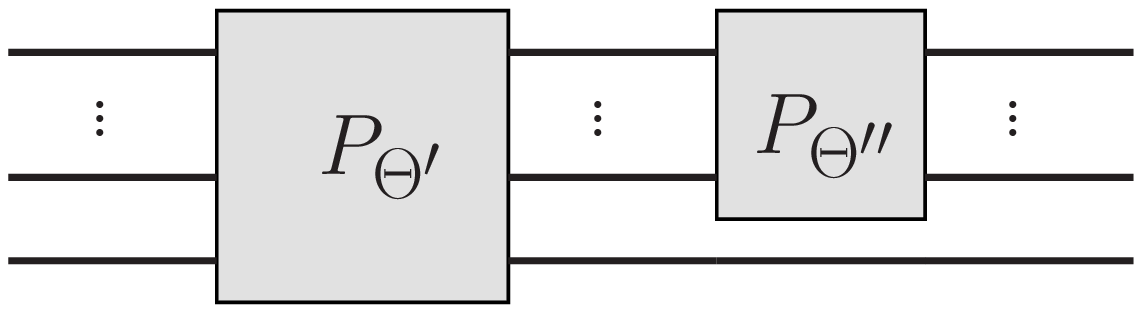}}
  = \parbox{5cm}{\includegraphics[width=5cm]{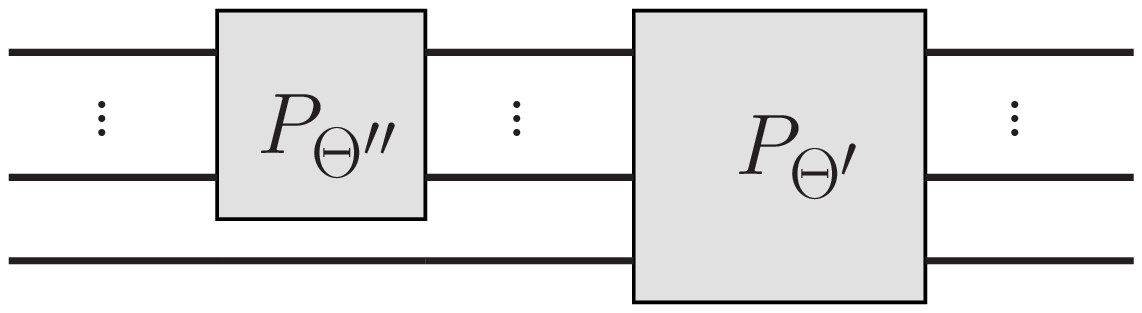}} \, .
\end{equation}
Therefore, we can insert factors of $\HYOp_{\YTab''}\otimes\eins$ to
the left and to the right of $\YOp_{\YTab'}$ in
Eq.~\eqref{eq:partial_trace_PTheta}, leading to
\begin{equation}
  \utr' \HYOp_\YTab 
  = (N+p-q) \frac{|\YTab'|}{|\YTab|} {\HYOp_{\YTab'}}^3
  = (N+p-q) \frac{|\YTab'|}{|\YTab|} \HYOp_{\YTab'} \, , 
\end{equation}
i.e.\ the Hermitian Young operators $\HYOp_\YTab$ fulfill the same
recursion relation as the conventional Young operators $\YOp_\YTab$,
cf.\ Eq.~\eqref{eq:Y_recursion}. Together with $\utr \HYOp_\YTab =
\utr \YOp_\YTab \ \forall\ \YTab \in \SYTx_2$ this proves
property~\ref{prop:dim} by induction.

Hermiticity of the operators $\HYOp_\YTab$ follows from
Corollary~\ref{corr:invariant_tensors_schur}, part~(ii): According to
the induction hypothesis the $\HYOp_{\YTab'}$ are Hermitian (and thus
project orthogonally), $\YOp_\YTab$ is an invariant tensor and a
projector onto an irreducible subspace, and due to property~(ii) $\utr
\HYOp_\YTab = \utr \YOp_\YTab \neq 0 \ \Rightarrow \ \HYOp_\YTab \neq
0$. Hence, $\HYOp_\YTab$ is proportional to the orthogonal projection
onto an irreducible invariant subspace. Moreover, since $\utr
\HYOp_\YTab = \utr \YOp_\YTab$ fixes the dimension of this subspace
the proportionality constant is unity, i.e.\ $\HYOp_\YTab^\dag =
\HYOp_\YTab$ and $\HYOp_\YTab^2=\HYOp_\YTab$, thereby also
establishing property (i) for the case $\YTab=\vYTab$.

I order to show property~\ref{prop:transversal} it only remains to
discuss the case $\YTab \neq \vYTab$, for which we have
\begin{equation}\label{eq:show_transversal}
  \HYOp_\YTab \HYOp_{\vYTab} 
  = (\HYOp_{\YTab'} \otimes \eins) \YOp_\YTab (\HYOp_{\YTab'} \otimes \eins)
    (\HYOp_{\vYTab'} \otimes \eins) \YOp_{\vYTab} 
    (\HYOp_{\vYTab'} \otimes \eins) \, .
\end{equation}
If $\YTab' \neq \vYTab'$ then $\HYOp_{\YTab'}\HYOp_{\vYTab'} = 0$ and
thus also $\HYOp_\YTab \HYOp_{\vYTab}$ vanishes. If $\YTab' = \vYTab'$
then $\YTab$ and $\vYTab$ have different shapes, and thus $\YOp_\YTab
\sigma \YOp_{\vYTab} = 0$ $\forall\ \sigma \in S_n$ according to
Lemma~\ref{lemma:inequivalent}. Since the terms between $\YOp_\YTab$
and $\YOp_{\vYTab}$ in \eqref{eq:show_transversal} are nothing but a
linear combination of permutations, $\HYOp_\YTab \HYOp_{\vYTab}$ also
vanishes in this case. 


The remaining property~\ref{prop:complete}, completeness, follows
straightforwardly from properties~\ref{prop:transversal} and
\ref{prop:dim}. To this end define
\begin{equation}
  P := \sum_{\YTab \in \SYTx_n} \HYOp_\YTab \, . 
\end{equation}
Then 
\begin{equation}
  P^2 \underset{\ref{prop:transversal}}{=} P 
  \quad \text{and} \quad 
  \utr P \underset{\ref{prop:dim}}{=} N^n \, , 
\end{equation}
which proves property~\ref{prop:complete} and thus concludes the proof of 
Theorem~\ref{thm:Hermitian_Young}.
\hspace*{1ex}\hfill\qed

As an illustration consider once more the two 5-box standard diagrams
$\tyoung{123,45}$ and $\tyoung{135,24}$ from
Sec.~\ref{sec:Young_operators}, whose conventional Young operators are
not transversal, see Eqs.~\eqref{eq:Littlewood_example} and
\eqref{eq:Littlewood_birdtracks}. Using the recursive construction of
Theorem~\ref{thm:Hermitian_Young}, the corresponding Hermitian Young
operators can be shown to be
\begin{equation}
\label{eq:Hermitian_Littlewood_birdtracks}
  \HYOp_{_\tyoung{123,45}} 
  = 2 \parbox{3.0cm}{\includegraphics[width=3.0cm]{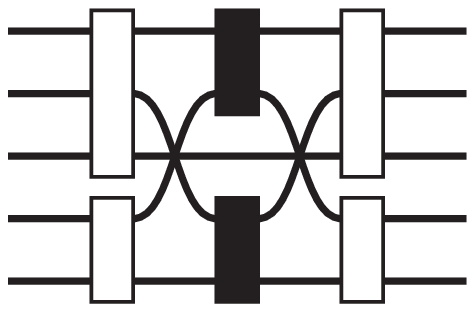}}
  \qquad \text{and} \qquad  
  \HYOp_{_\tyoung{135,24}} 
  = 2 
    \parbox{3.0cm}{\includegraphics[width=3.0cm]{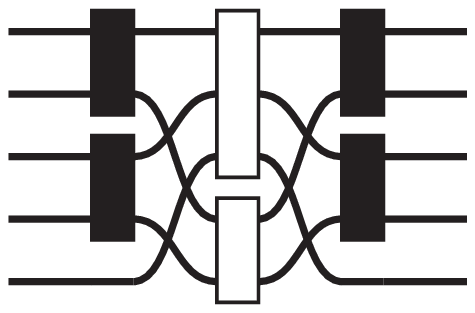}} \ . 
\end{equation} 
We give a step-by-step derivation in the Appendix. As
guaranteed by Theorem~\ref{thm:Hermitian_Young} these projectors are
not only Hermitian -- which is manifest from their birdtrack diagrams
being mirror symmetric -- but also transversal since
$\parbox{1cm}{\includegraphics[width=1cm]{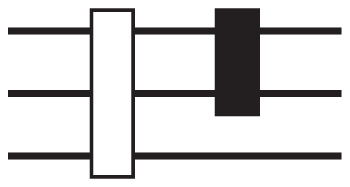}}=0$.

%
%

\section*{Acknowledgments}

We thank Johan Grönqvist and Jonas Lampart for helpful discussions. 
M.S. was supported by the Swedish Research Council, contract number 
621-2010-3326.

\section*{Appendix}

In order to get the Hermitian Young projection operator for the tableau
$\tyoung{123,45}$ we construct conventional and Hermitian Young
operators for its construction history, i.e.\ for the sequence
$\tyoung{12} \,,\ \tyoung{123} \,,\ \tyoung{123,4} \,,\
\tyoung{123,45} \,.$ The first few read 
\begin{equation} 
  \HYOp_{_\tyoung{12}} =  
  \YOp_{_\tyoung{12}} =  
  \parbox{1.5cm}{\includegraphics[width=1.5cm]{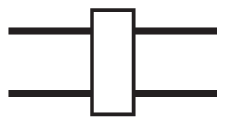}} \, , 
\end{equation}
\begin{equation} 
  \HYOp_{_\tyoung{123}} =  
  \YOp_{_\tyoung{123}} =  
  \parbox{1.5cm}{\includegraphics[width=1.5cm]{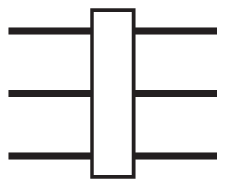}} \, , 
\end{equation}
\begin{equation} 
  \YOp_{_\tyoung{123,4}} =  \frac{3}{2} 
  \parbox{2.4cm}{\includegraphics[width=2.4cm]{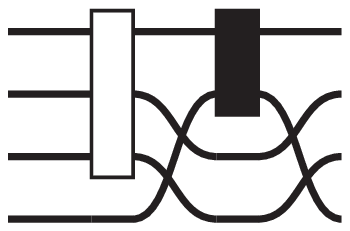}} \, , 
\end{equation}
\begin{equation} 
  \HYOp_{_\tyoung{123,4}} = \frac{3}{2} 
  \parbox{3.0cm}{\includegraphics[width=3.0cm]{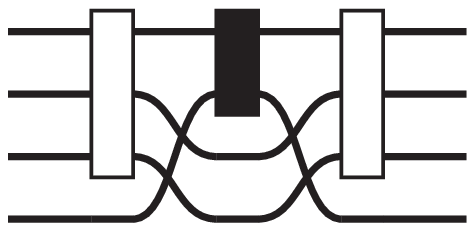}} \, , 
\end{equation}
\begin{equation} 
  \YOp_{_\tyoung{123,45}} = 2
  \parbox{2.4cm}{\includegraphics[width=2.4cm]{Pics/young123-45.eps}} \, .
\end{equation}
As there is only one subspace of $\left( \HYOp_{_\tyoung{123}} \otimes
  \eins^{\otimes 2} \right) V^{\otimes 5}$ carrying an irreducible
representation corresponding to the Young diagram
$\Yboxdim{6pt}\yng(3,2)$ we may in this case, according to
Corollary~\ref{corr:invariant_tensors_schur}~(ii), simplify the
construction as compared to that given in Eq.~\eqref{eq:Hermitian_Young_Def}:
It is not necessary to project $\YOp_{_\tyoung{123,45}}$ between
$\HYOp_{_\tyoung{123,4}} \otimes \eins$ but it suffices to calculate
\begin{equation}\label{eq:simpler_projection}
\begin{split}
  \HYOp_{_\tyoung{123,45}}  
  &= \left( \HYOp_{_\tyoung{123}} \otimes \eins^{\otimes 2} \right) 
     \YOp_{_\tyoung{123,45}}
     \left( \HYOp_{_\tyoung{123}} \otimes \eins^{\otimes 2} \right)\\[1ex]
  &= 2 \parbox{3.0cm}{
       \includegraphics[width=3.0cm]{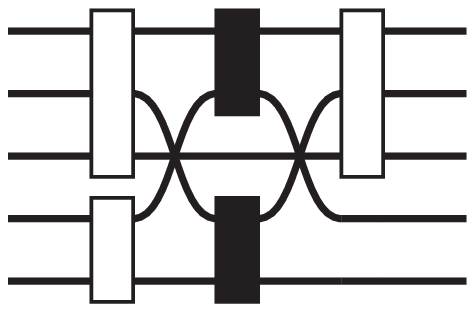}} \, .
\end{split}
\end{equation}
In order to make Hermiticity of this projector manifest, exchange the
positions of the two anti-symmetrizers, thereby thinking of the lines
as rubber bands which are pinned at the ends but which can pass though
each other. Keeping in mind that we can always rearrange the order in
which the lines enter a given symmetrizer we find
\begin{equation}\label{eq:twisted_rubber_bands}
  \parbox{3.0cm}{\includegraphics[width=3.0cm]{Pics/Hyoung123-45_step01.eps}}
  = \parbox{3.0cm}{\includegraphics[width=3.0cm]{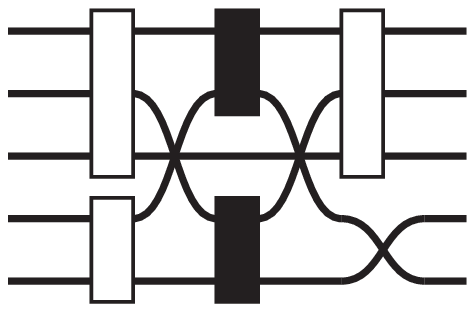}} 
    \, .
\end{equation}
Taking the average if these two expression we finally obtain 
\begin{equation} 
  \HYOp_{_\tyoung{123,45}} = 2
  \parbox{3.0cm}{\includegraphics[width=3.0cm]{Pics/Hyoung123-45.eps}} \, .
\end{equation}

The Hermitian Young operator for the tableau $\tyoung{135,24}$ can be
obtained analogously. This time we need to consider the
sequence $\tyoung{1,2} \,,\ \tyoung{13,2} \,,\ \tyoung{13,24} \,,\
\tyoung{135,24}\vphantom{^2} \,.$ The first few projectors read
\begin{equation} 
  \HYOp_{_\tyoung{1,2}} =  
  \YOp_{_\tyoung{1,2}} =  
  \parbox{1.5cm}{\includegraphics[width=1.5cm]{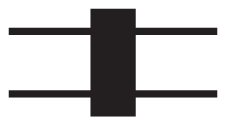}} \, ,  
\end{equation}
\begin{equation} 
  \YOp_{_\tyoung{13,2}} = \frac{4}{3} 
  \parbox{2.4cm}{\includegraphics[width=2.4cm]{Pics/young13-2.eps}} \, ,  
\end{equation}
\begin{equation} 
  \HYOp_{_\tyoung{13,2}} = \frac{4}{3} 
  \parbox{3.0cm}{\includegraphics[width=3.0cm]{Pics/Hyoung13-2.eps}} \, ,  
\end{equation}
\begin{equation} 
  \YOp_{_\tyoung{13,24}} =  \frac{4}{3} 
  \parbox{2.4cm}{\includegraphics[width=2.4cm]{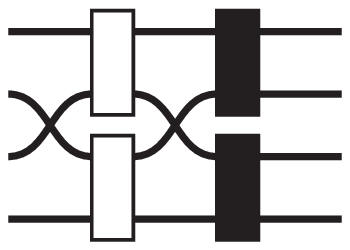}} \, .
\end{equation}
Similarly to the situation in Eq.~\eqref{eq:simpler_projection} there
is only one subspace of $\left( \HYOp_{_\tyoung{1,2}} \otimes
  \eins^{\otimes 2} \right) V^{\otimes 4}$ carrying an irreducible
representation corresponding to the Young diagram
$\Yboxdim{6pt}\yng(2,2)\vphantom{^2}$. Invoking again
Corollary~\ref{corr:invariant_tensors_schur}~(ii) we get
\begin{equation}
  \HYOp_{_\tyoung{13,24}} = \frac{4}{3}
  \parbox{3.0cm}{\includegraphics[width=3.0cm]{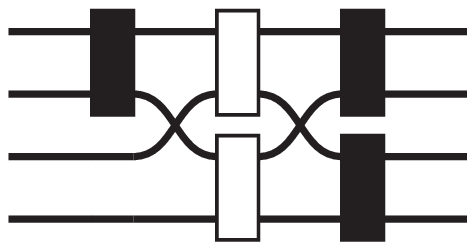}}  
  \, .
\end{equation}
Using a similar trick as in Eq.~\eqref{eq:twisted_rubber_bands}, this
time exchanging the two symmetrizers, the projector can be written in
manifestly Hermitian form,
\begin{equation} 
  \HYOp_{_\tyoung{13,24}} = \frac{4}{3} 
  \parbox{3.0cm}{\includegraphics[width=3.0cm]{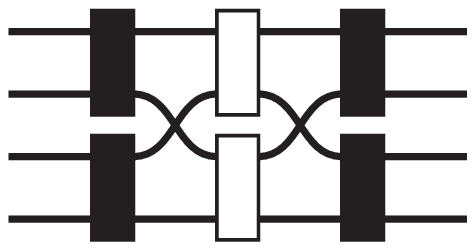}} \, .
\end{equation}
For the next step we have to sandwich  
\begin{equation} 
  \YOp_{_\tyoung{135,24}} = 2
  \parbox{2.4cm}{\includegraphics[width=2.4cm]{Pics/young135-24.eps}} 
\end{equation}
between two copies of $\HYOp_{_\tyoung{13,24}}\,$, 
\begin{equation}
  \HYOp_{_\tyoung{135,24}} = \frac{32}{9} 
  \parbox{7.0cm}{\includegraphics[width=7.0cm]{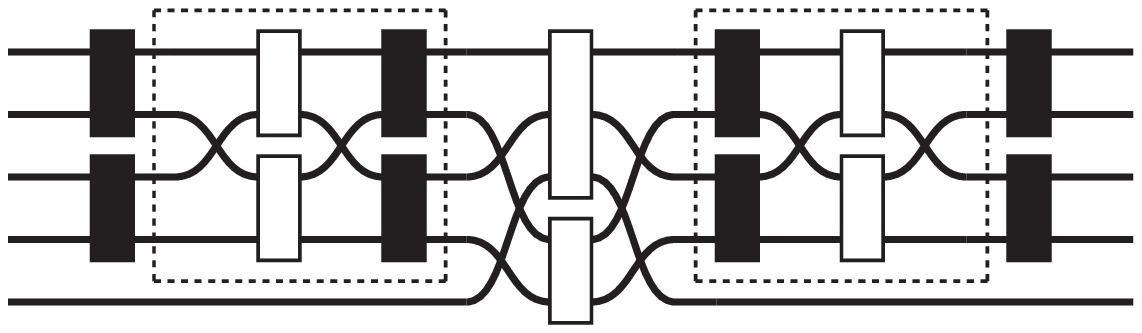}} 
  \, .
\end{equation}
In order to simplify this expression consider all lines passing
through the dotted boxes. On one side they connect to two
anti-symmetrizers, on the other side the connect to two
symmetrizers. In order for this connection to be non-zero each
anti-symmetrizer has to be connected to both symmetrizers. Thus, up to
re-ordering the lines entering a given symmetrizer, there is a unique
non-vanishing connection, i.e.\
\begin{equation} 
  \HYOp_{_\tyoung{135,24}} \ \propto \
  \parbox{3.0cm}{\includegraphics[width=3.0cm]{Pics/Hyoung135-24.eps}} 
\end{equation}
The normalization can be fixed by considering the square,
\begin{equation} 
  \left( 
  \parbox{3.0cm}{\includegraphics[width=3.0cm]{Pics/Hyoung135-24.eps}} 
  \right)^2 
  = \parbox{4.9cm}{\includegraphics[width=4.9cm]{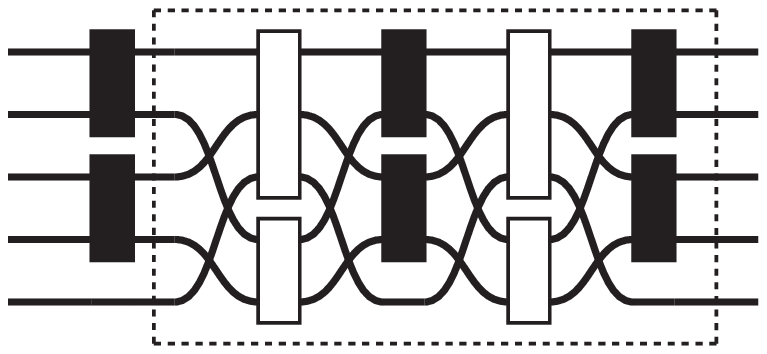}}
  \, .
\end{equation}
Here the dotted box is equal to
\begin{equation}
  \left( \frac{1}{2} \, \YOp_{_\tyoung{135,24}}\right)^2 
  = \frac{1}{4} \, \YOp_{_\tyoung{135,24}} \, , 
\end{equation}
and hence
\begin{equation} 
  \HYOp_{_\tyoung{135,24}} = 2
  \parbox{3.0cm}{\includegraphics[width=3.0cm]{Pics/Hyoung135-24.eps}} 
\end{equation}
is the desired projector.

\bibliographystyle{my_unsrt}
\bibliography{literatur} 

\end{document}